# Temporal sequences of spikes during practice code for time in a complex motor sequence

S. E. Palmer[1], M. H. Kao[2], B. D. Wright[2], and A. J. Doupe[2]

[1]Department of Organismal Biology and Anatomy, University of Chicago, Chicago, IL 60637.

[2]Departments of Physiology and Psychiatry, and Center for Integrative Neuroscience, University of California at San Francisco, San Francisco, CA 94158.

## Abstract

Practice of a complex motor gesture involves exploration of motor space to attain a better match to target output, but little is known about the neural code for such exploration. Here, we examine spiking in an area of the songbird brain known to contribute to modification of song output. We find that neurons in the outflow nucleus of a specialized basal ganglia-thalamocortical circuit, the lateral magnocellular nucleus of the anterior nidopallium (LMAN), code for time in the motor gesture (song) both during singing directed to a female bird (performance) and when the bird sings alone (practice). Using mutual information to quantify the correlation between temporal sequences of spikes and time in song, we find that different symbols code for time in the two singing states. While isolated spikes code for particular parts of song during performance, extended strings of spiking and silence, particularly burst events, code for time in song during practice. This temporal coding during practice can be as precise as isolated spiking during performance to a female, supporting the hypothesis that neurons in LMAN actively sample motor space, guiding song modification at local instances in time.

The vocalizations of songbirds represent a classic learned sensorimotor skill: initially immature, variable sounds are gradually refined through extensive rehearsal during which auditory feedback is used to guide learning as well as adult modification of song. A variety of evidence suggests that a cortico-basal ganglia circuit dedicated to song is crucial for this trial and error learning: lesions or inactivation of the 'cortical' outflow nucleus of this circuit, the lateral magnocellular nucleus of the anterior nidopallium (LMAN; see **Fig. 1a**), cause an abrupt reduction in song variability and a failure to progress towards a good match to the target song[1-3]. These findings suggested the hypothesis that LMAN actively injects "noise" into the motor system to generate vocal variability and facilitate motor exploration for trial-and-error reinforcement learning[3-6].

It has been directly demonstrated that even subtle, trial-by-trial variations in the acoustic structure of adult song can be used to drive rapid, adaptive changes in song[7,8]. For example, if a loud burst of white noise is played each time a bird sings a particular song element ('syllable') at a high pitch, but not at a lower pitch, the bird will gradually change the pitch of the targeted syllable in order to avoid the aversive white noise. These data show that behavioral variability ('motor exploration') in song can be used to enable adult song plasticity in a negative reinforcement-learning paradigm. Thus, the bursting output of LMAN neurons during singing has been hypothesized to carry some instructive signal that locally perturbs song output. Here, we explore the question of what parts of LMAN firing during singing are most reliably locked to time in song, in an attempt to discover what temporal patterns contain the code for this song variation.

In addition, both song variability and neural variability are strongly modulated by social context. When males sing courtship song directed at females, both the acoustic structure of individual syllables and syllable sequence are more stereotyped than when males sing alone ('undirected' song[5,9-11]). These context-dependent changes in song variability are accompanied by striking changes in the singing-related activity of LMAN neurons: LMAN neurons exhibit more variable spike timing and more frequent burst firing during undirected singing (e.g., **Fig. 1b**; [12,13]). Spike timing variability in LMAN and song variability are even higher in juvenile birds actively engaged in sensorimotor learning[3,6,10,14-16]. Strikingly, the same lesions or inactivation of LMAN that eliminate song plasticity also reduce song variability during undirected song to the level observed during directed singing[11,17,18]. In addition, manipulations of the AFP circuit that specifically eliminate LMAN bursts (but not single spike firing) also eliminate the bird's ability to change song in response to altered auditory feedback[19]. The differential effectiveness of burst firing versus single spikes has been demonstrated in a variety of *in vivo* and *in vitro* settings (see [20] and [21] for reviews). Together, these results suggest that the variable LMAN bursting typical during undirected singing drives song variability and plasticity. Moreover, they support the idea that undirected song reflects a "practice" state, in which behavioral variability enables maintenance and/or optimization of song, while directed song reflects a "performance" state in which a male sings his current best version of song[17,22-25].

While previous theories suggested that LMAN injects random variability in the motor pathway that enables song change[3,6,26], recent data suggest that LMAN firing can target particular local variations in song output. Based on these findings, we hypothesize that the variable LMAN bursting during undirected singing is not simply noise but carries information about song. To quantify this, we compute the mutual information[27-29] between temporal patterns of spiking in LMAN and time in song. We find that these patterns carry information about song, particularly during undirected singing, and that the neural code is fundamentally different depending on behavioral context. The importance of temporal pattern coding for song has also been quantified in the motor nucleus of the song system (RA), to which LMAN neurons project[30].

## Results

To assess whether the output from the anterior forebrain pathway (LMAN) injects a noise-like signal to the motor system, we analyze data from single LMAN neurons in adult male zebra finches during interleaved bouts of undirected singing and courtship singing. Song is an extended motor sequence that maintains a stereotyped structure from rendition to rendition in adult male birds. Some of this variation is in acoustic features of the song, and some is in the length of pauses between syllables as well as the length of the syllables themselves. To remove the second source of variation, songs and spike trains are time-warped to remove the slight differences in length of pauses and syllables across renditions so that we can compare the code for the same song across renditions (see Methods). There are an infinite number of ways to quantify the relationship between spiking and song features, but we can use time in song as a proxy for the whole host of acoustic features present at that moment in time. Then, to quantify the correlation between spiking and song, we compute the mutual information between spiking patterns and time in song. We first assess the temporal precision and information content of single spikes, which to the eye appear more noisy during undirected singing.

*Single spike information is higher during directed than undirected singing*
We begin by quantifying the mutual information between time in song and the arrival time of single spikes in both behavioral contexts. The mutual information quantifies the generalized correlation between two variables, here probability of spiking and time in song. Intuitively, this measures how precisely firing rate modulations are locked to a particular part of song. Since song is consistent from rendition to rendition in adult zebra finches, time in song can be used as a proxy for the particular motor gesture at that time[31-34].

As reported previously, visual inspection of the spike trains for single LMAN neurons show striking differences in the reliability and precision of spikes across repeated renditions of song depending on behavioral context, even though the song is very similar, acoustically, across conditions[12,13]. For example, in Figure 1b, left, when a male sings songs directed to a female, spikes align across song renditions, and the average firing rate for this cell (red trace) is strongly peaked across repeated renditions of a stereotyped sequence of song elements ("motif"). The targeting of spikes to particular parts of song may reflect a refinement of the brain's firing during the task, akin to behavior-locked sharpening of response during learning[35,36]. When we compute the temporal jitter in the arrival times of the spikes, we find that during directed singing, spikes from this site are timed with a resolution of approximately 2.6 ± 1.6 ms, as calculated by the standard deviation in timing of spikes during peaks in firing probability (see Methods). In contrast, for the same neuron, the spikes recorded during undirected song are more variable across renditions and have a temporal jitter of 3.9 ± 2.2 ms (**Fig. 1b**, right). Indeed, across the population of LMAN neurons, single spikes have an average jitter of 4.5 ± 3.1 ms during undirected singing versus 2.8 ± 0.6 ms during directed singing, indicating that the timing of individual undirected spikes is less precisely locked to a particular time in song. On average however, the firing pattern during undirected singing has a similar shape to that during directed singing (**Fig. 1b**, blue trace, with directed pattern overlaid in light red for comparison).

To further quantify these observations, we use information theory to compute a single number that measures the generalized correlation between the arrival time of single spikes and time in song. This reveals where in song spiking output from LMAN is most keenly focused. Computing the mutual information in this way is usually performed on data recorded during repeated presentations of a sensory stimulus[28]. Here, we are using the same formalism to compute the information about time during a repeated motor sequence. In this way, we measure how LMAN spiking correlates with output. To do so, we compute the average log signal-to-noise ratio of the firing rate, .

$$I = \frac{1}{T} \int_0^T dt \left(\frac{r(t)}{\bar{r}}\right) \log_2 \left(\frac{r(t)}{\bar{r}}\right), \tag{0}$$

where $r(t)$ is the trial-averaged firing rate as a function of time during song, for each recorded unit, akin to the peristimulus time histogram, but here labeled the peri-song time histogram (PSTH). Observing spikes from a neuron with a sharply peaked PSTH would give good information about which parts of song were currently being sung. By contrast, a spike from a neuron with a flat PSTH gives no information (no discriminability) about time in song. Another way to think about the mutual information computed here is that it quantifies how reliably peaked and inhomogeneous the firing of the neuron is during song. Observing a spike from a neuron with a high information content means that one can infer with high reliability where you are in song. For the bird, this means that these particular parts of song are potentially being targeted for modification or reinforcement.

We compute information about time in song for all sites, with a time resolution of 2ms, and find that there is significantly more information in single spikes during directed song than during undirected song at each site, as shown in **Figure 1c** ($p < 0.001$), consistent with the greater temporal precision of single spikes in this context. These results indicate that on an individual trial, single spikes convey more information about time in song in the directed condition (more bits per spike) than in the undirected condition. In addition, higher firing rates during undirected singing (more spikes per second) do not compensate for the lower information content per spike. When context-dependent firing rate differences are taken into account, the single spike information is still significantly lower during undirected song (**Fig. 1d** shows information in bits/sec).

*Pattern information is similar during undirected and directed singing*
Lower single spike information during undirected singing could arise if the same precise firing observed during directed song is obscured by accompanying noisy burst firing (**Fig. 2a, parts i-ii**). This would support the hypothesis that LMAN activity generates a noise signal for motor exploration during a practice state. Alternatively, the bursts observed during undirected singing could reliably pick out particular times in song depending on their internal spiking structure, and only appear noisy when different temporal patterns are lumped together into a single category (e.g., bursts; **Fig. 2a, part iii**). Indeed, the single spike information measure assumes that all successive spikes are independent events, and thus explicitly neglects any information that might be present in temporally extended patterns of firing.

To analyze information from sequences of spikes, we extract short temporal strings (10ms) of activity from the binned spike data. As the example in **Figure 2b** illustrates, individual spikes are denoted by a '1' and silences by a '0' in 2ms bins. We then obtain the complete distribution of 5-bit patterns observed in these 10ms windows taken from all parts of song. The probability distribution of patterns for directed and undirected spiking is shown in **Figure 2c**, with the patterns grouped according to their total spike count in the 10 ms window. This count can range from 0-5, with 5 representing a spike in every 2 ms time bin. The higher total spike count in the undirected condition reflects both the observed bursts present in these spike trains and the higher firing rate in undirected singing. The probability of observing a spiking pattern with more than 2 spikes in the full 10ms window is much smaller during directed song than undirected song.

The mutual information contained in temporal patterns about time in song computed using the direct method[27]; see Methods and Supplementary Information) is defined by the overall entropy of patterns (the 'vocabulary' of LMAN spiking in a particular condition) minus the variability in spiking output at a particular time during song (the 'noise' entropy):

$$I(patterns; t) = S(patterns) - \langle S(patterns|t) \rangle_t.$$
(0)

This pattern information is plotted in **Figures 2d-g**. To compute the entropy, a measure of variability, of patterns at a particular time (the second term in the sum), we need to have enough samples at that time to estimate the full $2^{\wedge}nbins$ parameters. Care is taken to ensure that sampling bias does not affect this measurement including choosing sites depending on how many samples we have at each instance of time and comparing to other entropy estimation procedures that show the best performance in the not-strongly-sampled regime (supplementary information). When temporal patterns of spiking are taken into account, information about time in song increases in both contexts, but much more so in the undirected condition (**Figures 2d-e**). A particular pattern might be very informative, but also very rare. To de-emphasize this contribution to the total pattern information, we measure the mutual information in bits per second. Rare patterns do not make significant contributions to the information rate. The relative increase in information rates obtained for patterns versus single spikes is plotted in **Figure 2f**, as a function of the number of 2ms bins we observe (the length of the temporal pattern). As we increase the number of bins in our temporal strings, we are restricted to sites with higher numbers of repeated song renditions (illustrated in **Supplementary Fig. 2**), and at 7 or 8 bins, we only have a few sites for which we can reliably measure information values. The ratio of pattern information to single spike information reveals a larger increase in undirected singing, and this difference is significant for 5 and 6 bin strings. The absolute pattern information between conditions (shown in **Supplementary Fig. 3)** equalizes at N bins = 6. During directed singing, single spikes and their timing precision carry most of the information about time in song. In contrast, during undirected singing song information is largely carried in the temporal patterns of spikes and silences. In this behavioral context, there must be some patterns that exhibit greater precision in their timing than that of the average single spike. While the overall information rate is still significantly greater during directed singing, context-dependent differences in the information rate are reduced when we take into account temporal patterns (e.g. compare **Fig. 1d** to **Fig. 2g**).

*Pattern information is conveyed by different spiking sequences in the two contexts*

By examining the precision of particular patterns of spikes in time, we are able to extract those patterns that carry information about song during undirected singing. These specific patterns are obscured in the full spike raster, where all patterns that occur are plotted together (see **Fig. 2a, part iii**). **Figures 3a-b** show two firing patterns during undirected singing from two different LMAN neurons that are much more tightly locked to song than the average single spike. Sharp peaks in the pattern probability (indicated by arrows in middle and bottom panels) occur at points in song where the single spike PSTH (top panel) was only weakly modulated above the mean.

Given that we know that patterns carry song information and that some patterns are precisely timed with respect to song, we next examine what characterizes these informative patterns. Is the total pattern information distributed uniformly across all possible patterns or is it clustered around particular patterns?

The full pattern information can be decomposed into a sum over all $2^5$ possible strings of spikes and silence,

$$I_{patterns} = \sum_{i} I(pattern_i) P(pattern_i),$$

(1)

where the information in a particular pattern is computed using that particular pattern's PSTH using Equation 1. To examine which patterns carry information about song, we group the observed patterns according to how many total spikes are observed in the 10 ms window. For example, we can compute how much information all of the two-spike patterns carry about song by collecting the terms in the sum in Eq. 3 that correspond to the 10 unique 2-spike patterns. There is just one zero-spike word, the pattern with all zeros, 5 one-spike words in which the single spike occurs in one of each of the 5 bins, $\binom{N}{2}$ two-spike words, and so on. Grouping patterns in this way, we plot the total pattern information, decomposed according to K-spikes in the pattern in **Figures 3c-d** for directed and undirected song. Single-spike patterns account for most of the information observed during directed song (70 ± 4%). These results are shown for the average over all sites, but each individual cell also displayed the same trend. The greater precision of single spike timing during directed song gives rise to larger information in these patterns than in undirected song. In contrast, pattern information during undirected song is dominated by 2- and 3-spike strings, which are primarily bursts (59 ± 5% of the total information is from 2 and 3 spike patterns, all of which are classified as bursts when defined by an instantaneous firing rate > 200 spikes/s). **Figure 3d** shows that bursting during undirected song is not just noise; rather, bursts are the main contributors to information about time in song. This implies that the coding scheme used by an individual neuron during these two behavioral contexts is quite different: depending on context, a different set of temporal patterns is used (**Fig. 2c**) and these patterns contain different amounts of information (**Figs. 3c-d**).

*Coding is more rate-like during directed song and more temporal during undirected song*

To investigate directly whether the same neuron can use different coding schemes depending on social context, we eliminate timing information within the window so that we can compare the full pattern information to simple count information (**Fig. 4a**). To examine which patterns carry more than just spike count information, we compare pattern information to count information for a given number, K, of spikes in the window in **Figures. 4b-c**. In the directed case, about half of the pattern information can be summarized by count; simply noting whether 0, 1 or 2 spikes occurred in a 10ms window yields more than half (64%) of the total information about time in song. The greatest difference between full pattern information and count is explained by the timing precision of the first spike in a window, for patterns containing a single spike. In contrast, count information is consistently low in the undirected condition, even for high-count events, and accounts for only 39% of the complete pattern information. Moreover, higher firing rate during undirected singing does not yield more information in spike count.

To the extent to which pattern information can be summarized by counting spikes, the encoding is rate-like with a time resolution of the pattern length (here, 10ms). **Figure 4d** plots the pattern information versus count information for each neuron, and shows that the data roughly cluster into two groups according to social context. Information during directed singing (red circles) trends along the rate-coding line (count=pattern), while pattern information in undirected singing (blue circles) outstrips the very small count information. The change in the relative amount of count and pattern information with behavioral context indicates that single neurons switch between more rate-like coding during directed song and temporal coding during undirected song.

The relative ratio of count to full pattern information in the two contexts is shown in **Figure 4e**. The gray bars show the fraction of total pattern information that can be summarized by counting the number of spikes in the 10ms window. Count information is a substantially larger fraction of the total pattern information during directed singing. We can add additional temporal information to the spike count to see what kinds of temporal coding contribute to the observed pattern information. If we keep track of both count and the timing of the first spike in the window (a measure of temporal precision in the initiation of firing in the window), nearly one-third of the total pattern information during undirected singing is not recovered (**Fig. 4e**, white region). This remaining information is the pure temporal sequence part of the code. This means that the detailed timing of spikes within bursts contributes substantially to the code for time in song during undirected singing.

*A Poisson model does not explain temporal coding during practice*

What elements of the firing in LMAN give rise to the temporal code we observe during practice? Can the time-varying firing rate measure at 2ms resolution account for all of the observed information or do trial-to-trial correlations add significantly to the code? To address these questions, we compute the same mutual information quantities between time in song and spikes generated by a time-varying Poisson process with a rate matched to the real neurons, with varying temporal resolution. A Poisson model with a rate varying at the timescale of our temporal pattern window (10ms) does not contain any information beyond the spike count in the window (data not shown). We can pin the rate to follow the same time course as the real neurons (2ms resolution) as well and measure the information in temporal patterns from these model responses. We find that although some information about time in song is reproduced, significant amounts of information remain, particularly during undirected firing (data not shown). Overall, the model fails to capture the details of the coding switch between performance and practice singing.

*Spikes within bursts code for time in song synergistically*
How can pattern information arise in undirected singing from relatively uninformative single spikes?  The answer clearly has to do with the fact that subsequent spikes and silences refine an estimate of time in song, adding up independently at least, and admitting the potential for synergy in temporal sequences of activity.  **Figure 5a** illustrates how subsequent spikes in a burst carry additional information about time in song, depending on exactly when in the window they occur. During undirected song, this typical LMAN neuron has several 2-spike patterns in its vocabulary, whose probability of occurrence peaks during different parts of song.  The particular time in song coded by the pattern depends on the precise temporal pattern of spiking and silence. For example, the pattern "01001" peaks at ~150ms in the motif, while "10001" peaks at a completely different time in song (~450ms).  Different patterns sometimes also have peak firing rates at different times in song during directed singing, but mostly code for the same part of song redundantly (see **Supplementary Fig. 4** in the Supplementary Information).  These results indicate that the larger repertoire of temporal patterns used to convey information during undirected singing is neither an artifact of a higher firing rate nor a switch to random bursting, but rather is used to encode different times in song.

To dissect this more carefully, we quantify how spikes and silences combine to create the observed encoding.  We compute the information from the full string of spikes and silences in an observed pattern, minus the sum of the contributions from each spike or silence individually,

$$I_{synergy} = I(\{\sigma_i\}; t) - \sum_i I(\sigma_i; t), \tag{1}$$

where $\sigma$ is a binary variable for a spike $(\sigma = 1)$ or silence $(\sigma = 0)$ and $\{\sigma_i\}$ is the 5-bit string of spikes and silences in the temporal pattern.  If this quantity is positive, the pattern synergistically codes for time in song, meaning that spikes in the pattern add information supra-linearly.  If negative, the spikes and silences add redundantly, meaning that additional spikes in the burst code for the same part of song.  If spikes and silences independently code for time in song, their synergy is zero.  We compute the synergy for each pattern, then group patterns as in **Figures 2-3**, and plot this value as a function of the number (K) of spikes in the pattern (**Figs. 5b-c**).  Most strikingly, we see significant synergy in patterns of spiking with more than 1 spike per pattern, particularly 2 spikes and greater.  This clearly indicates that bursting, especially in the undirected song, carries information about song in an intriguing way not often seen in neural coding. Indeed, if LMAN bursts were generated by a cellular mechanism intrinsic to the cell, we would expect subsequent spikes in a burst to form a redundant code for time in song, since their pattern was completely determined by burst initiation.  LMAN neurons are not intrinsic bursters (ref), however, so the bursts that we observe are the product of input activity and recurrent circuit dynamics and may contain significant tunable temporal correlation structure.

Bursts are not unitary events that can be combined into a single category or even subdivided by the number of spikes within the burst.  Rather, the particular pattern of spikes within a burst conveys information about time in song, and does so synergistically.  In the directed condition, bursts consist almost exclusively of 2 spikes, and while they display synergy, the proportion of pattern information that arises from synergy is less than that observed during undirected singing for both N bins = 5 and 6 (**Figs. 6a-b**).  This trend continues through N bins = 7 (**Fig. 6c**).  During directed singing, we also observe significant information in strings of silences (0-spike patterns) in the directed condition.  This is surprising at first, since single silences carry very little information about position in song, but we can understand this as simply the complement to the precision of single spike timing.

Overall, our results directly challenge the notion that LMAN is a simple noise generator and support recent data suggestive that LMAN activity locally modulates acoustic features of song. Bursts carry the majority of information about time in undirected song via a truly temporal pattern code. We observe that behavioral context elicits a state change in LMAN firing, in which the code for time in song changes from a primarily single spike-timing code during directed singing, to a code dominated by temporal sequence encoding during undirected song.

## Discussion

Our analysis reveals that the precise timing of spikes and silences of single LMAN neurons carries information about time in song, and that the neural code is fundamentally different depending on behavioral context. While it has been hypothesized that the variable firing patterns in LMAN inject noise into the bird's motor system to facilitate exploration for trial-and-error reinforcement learning [3,6,26], we find that LMAN firing patterns come in many forms, each of which can reliably indicate particular times in song depending on their internal spiking structure (e.g., see **Figures 3a, 3b,** and **5a**), more consistent with recent experimental observations of the local effects LMAN stimulation can have on song pitch and loudness (need to ref here and check). When these patterns are lumped together, however, they appear to be noisy. Moreover, the diversity of patterns used to encode information about song changes with behavioral context: information is carried predominantly by precisely timed bursts during undirected singing, a putative practice state, versus the precise timing of single spikes during performance directed at a female.

Changes in the firing mode in individual cells and entire circuits can shape how signals are transformed within and received downstream of that brain area. Bursts, in particular, are thought to represent an important mode of neuronal signaling. In comparison with single spikes, bursts have been shown to transmit stimulus information in a way that is distinct from single spikes both in the neocortex and in the midbrain of weakly electric fish[20,37-40], and to enhance the reliability of information transfer in the hippocampus and in the lateral geniculate nucleus (LGN)[21,41,42]. In songbirds, a variety of evidence indicates that burst firing in LMAN may be particularly important for driving changes in song. First, manipulations of the AFP circuit that specifically eliminate LMAN bursts (but not single spike firing) also eliminate the bird's ability to change song in response to altered auditory feedback[19]. Indeed it is harder to perturb song during directed singing than during undirected singing[43], perhaps because isolated spikes in LMAN do not facilitate downstream plasticity as well as bursting. The suppression of burst firing and the increased reliability and precision of spike timing in LMAN that is elicited by the presence of a female is reminiscent of the observed decline in neural variation with stimulus onset in many areas of the mammalian cortex[44].

The diversity of burst patterns deployed during undirected song may help guide downstream motor behavior toward a targeted, moment-by-moment exploration of song space through multiple mechanisms. First, particular burst patterns could specify the specific parts of song to modify as well as how to change their acoustic features by altering the firing rates of neurons in the song motor cortex (RA, robust nucleus of the arcopallium, **Fig. 1a**), which have been shown to correlate with the pitch, amplitude, and local entropy of individual learned song elements[15,45,46]. Indeed, acute alterations of LMAN firing via microstimulation can drive systematic changes in acoustic features of song[13]. Second, the specific temporal patterns within bursts could have different downstream effects via nonlinear dendritic[47] or network[48] interactions. Third, burst firing in LMAN could facilitate plasticity via NMDA receptor activation[49,50] and increased calcium influx in RA neurons. LMAN might also impact the sequence of syllables sung more indirectly via projections onto HVC[51].

Bursting underlies the coding switch with behavioral state in this system, which is the analog to the outflow of mammalian basal ganglia circuitry. While bursting is often thought of as a pathological signal in basal ganglia circuits in disease states, such as the increased bursting observed in Parkinson's disease[52], with therapies such as deep brain stimulation deployed to suppress it[53], here we have shown that it might play a role in shaping normal task-specific behavior, like novelty-driven exploration[54]. Additionally, context-triggered switches in coding like the one seen here may be a general feature of basal ganglia circuits, which are well known to show sensitivity to contextual cues such as reward[55,56].

More broadly, this work addresses how reinforcement learning may be instantiated in any motor system. In contrast to models of reinforcement learning that use random noise to generate behavioral variability in order to thoroughly explore task space to achieve optimal behavior, avoiding the pitfalls of local minima[57], we suggest that the nervous system may actively sample motor space in a targeted manner. We find that the precise timing of spikes in temporally extended patterns can reliably signal particular times in a task and potentially direct motor exploration towards the target. Our results support theories of sensorimotor learning that permit active sampling of motor space. Such 'active learning' models are more efficient, displaying an exponential improvement in the number of samples needed to reach target generalization error as compared to random or batch-learning models, and are gaining popularity in the machine learning community[58-61]. Our results speak to their general applicability to neural systems.

## Acknowledgements
We thank Gašper Tkačik and David Schwab for useful conversations, and Philip Sabes and members of the Doupe laboratory for feedback on the manuscript. This work was supported by NIH grants MH55987 and MH78824 (to A.J.D.), by the Swartz Foundation, and in part by NSF grant 1066293 and the hospitality of the Aspen Center for Physics.

## Author contributions
S.E.P. and B.D.W. conceived of the analysis. M.H.K. and A.J.D. designed and M.H.K performed the experiments. S.E.P. performed the data analysis and modeling. S.E.P, M.H.K and A.J.D. wrote the paper.

## Methods
*Recordings*
Single unit recordings were made from area LMAN in 28 sites from 9 adult male zebra finches, as previously reported[13]. Spiking activity was measured during bouts of singing both to a female bird and when the male bird was alone, as well as when the male bird was silent.

*Song warping*
Songs were segmented and aligned using linear warping, as previously described[13]. This warping allows us to align spike trains across renditions, but does not remove variability in spectral features of song from rendition to rendition.

*Defining spike patterns*

Time was binned in 0.5, 1, 2, 3, and 4 ms bins, and the particular bin size chosen did not affect our results. Data shown in the figures are for 2 ms bins.   Patterns of spikes were classified as bursts when the instantaneous firing rate exceeded 200 spikes/s at any time within the window, i.e. when any two spikes were separated by less than two bins of silence.

*Information in spikes, spike count, and spike patterns*

To quantify information contained in spikes, we computed the information about time in song using the direct method[27].   For single spikes, the information about time in song is related to the average modulation of the firing rate divided by the mean rate over the entire trial.   To proceed beyond single spike information, we characterize the average entropy in temporal patterns as a function of time in song.  This so-called noise entropy is subtracted from the total entropy of patterns (vocabulary size minus noise), yielding Eq. 2.   In a similar fashion, the total count in the length T window can replace pattern, and this entropy can be computed both overall and as a function of time in song.  Other coding schemes can be compared to the total pattern information, as partitions of the patterns into different bins (count and time of the first spike in the window, for example).   All information measures are prone to sampling biases.   To correct for this, we use quadratic extrapolation to the infinite data size limit [27,62], using 50 bootstrap samples of the data each at of the following data fractions: 0.95, 0.90, 0.85, 0.80, 0.75, 0.70, 0.60 and 0.50.   Error bars on information quantities were assessed by taking the standard deviation of the 50% data samples and dividing by sqrt(2). Information is also computed for shuffled data, where all labels have been randomly reassigned.   Any residual information in this shuffled data is due solely to data limitations.  We exclude data when the shuffle information is not within error bars of zero.  We also exclude sites for which the total number of observed patterns of spiking and silence was greater than the number of trials.   These values are illustrated in **Supplementary Figure 2.**

*Statistical tests*

The significance of the observed differences in information measures across the two conditions was assessed using a Student's t-test for the total population data, and a paired Student's t-test for the sites in which both directed and undirected singing data were collected.   Significance values were denoted with asterisks:   p < 0.01 *, p < 0.001 **, p < 0.0001 ***

*Synergy*

The difference between the total pattern information and the average information carried by individual spikes and silences in the pattern was calculated as in Eq. 4, as proposed by Bialek and coworkers[28].  We also compute what fraction of the total pattern information is due to synergy, by summing over synergy values for each pattern, weighted by that pattern's probability.

Figure 1

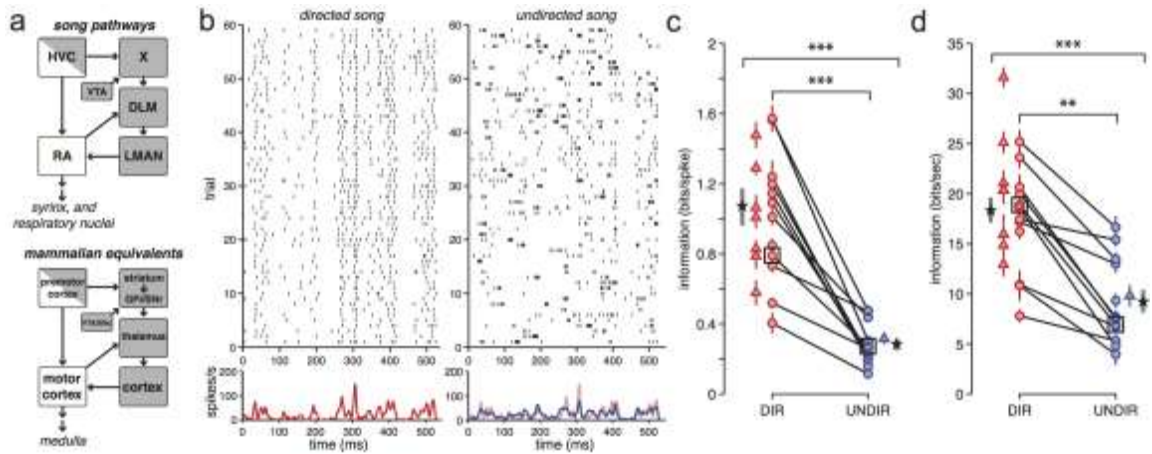

**Figure 1: Single spike information about time in song is higher during directed song.** (**a**) Schematic diagram of the main brain areas involved in song learning and production and corresponding areas in the mammalian brain. (GPi, internal segment of the globus pallidus; VTA, ventral tegmental area; HVC, used as a proper name; SNr, substantia nigra pars reticulata; LMAN, RA, and X are defined in the main text.) (**b**) Spike rasters (top) and corresponding averaged firing rates smoothed with a Gaussian kernel with standard deviation (SD) = 2ms [peri-song time histograms (PSTHs), bottom) for one LMAN neuron during directed (left) and undirected (right) singing show increased firing rate, more bursts, and more apparent noise during undirected singing. (**c**) Information from single spike arrival times in LMAN during directed song (red) and undirected song (blue), measured using Eq.1. Data are shown with a bin size, $\Delta t = 2\text{ms}$. Lines connect data from single neurons with recordings in both conditions. Black boxes indicate the single spike information for the neurons shown in **b**. Triangles indicate neurons from which recordings were made during only one context. Stars and gray bars indicate the mean ± standard error (SEM) across all sites. Here, and throughout, red denotes the directed singing condition, blue undirected. (**d**) Information rate from single spike arrival times, plotted in units of bits/sec. (*** indicates p < 0.0001, ** p < 0.001, * p < 0.01, paired Student's t-test for comparisons of sites measured in both conditions, unpaired Student's t-test for population mean comparison, combining both paired and unpaired data).

Figure 2

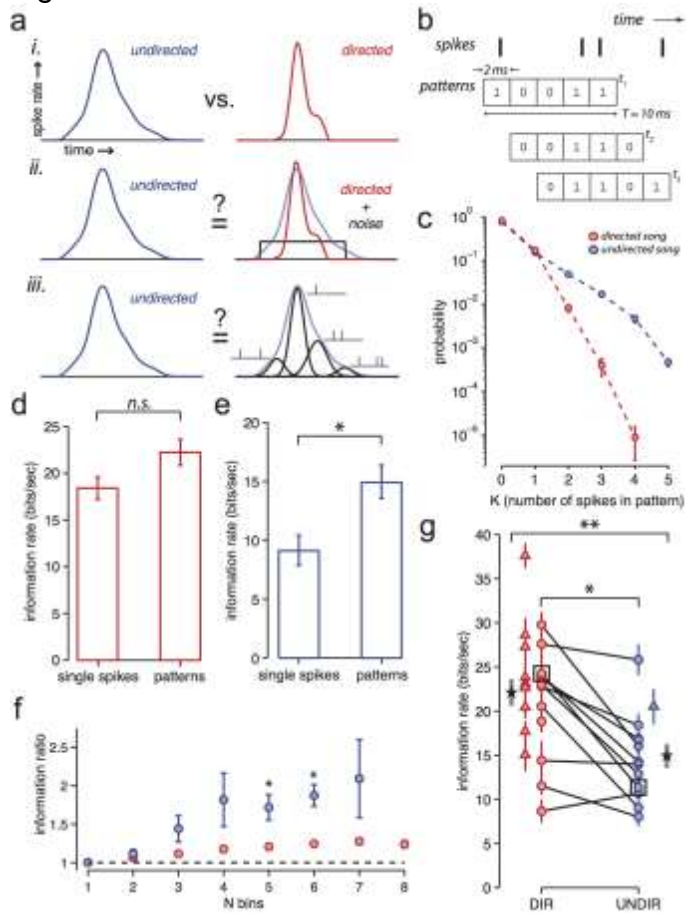

**Figure 2: Information from sequences of spikes and silence are comparable during directed and undirected song.** (**a**) We illustrate two hypotheses about the relationship between firing across behavioral conditions: First, spiking during undirected song consists of isolated spikes with the same firing pattern as in directed song (i), but against a background of noisy bursts (ii). Second, bursts are precisely timed to song, but to different parts of song depending on the particular temporal sequence of spikes in the burst (iii). (**b**) An illustration of how we define our spiking symbols for temporal pattern calculations. Spike trains are binned at 2ms resolution. Patterns are defined in a 10ms window. Full 5-bit patterns of spiking (1) and silence (0) are retained for computing pattern information. (**c**) The probability of observed patterns, grouped by the number of spikes in the 10ms window ('spike count'), for directed (red) and undirected (blue) song. (**d**) and (**e**) Pattern information compared to single spike information averaged across all sites during directed and undirected singing. (**f**) The ratio of pattern information to single spike information reveals a larger increase in undirected singing as a function of the length of the temporal pattern. (**g**) A small but significant difference in pattern information is observed across recording sites during directed (red) versus undirected (blue) singing. Black boxes indicate the pattern information for the site shown in **Figure 1b**. p-values are indicated by asterisks as in **Figure 1**.

Figure 3

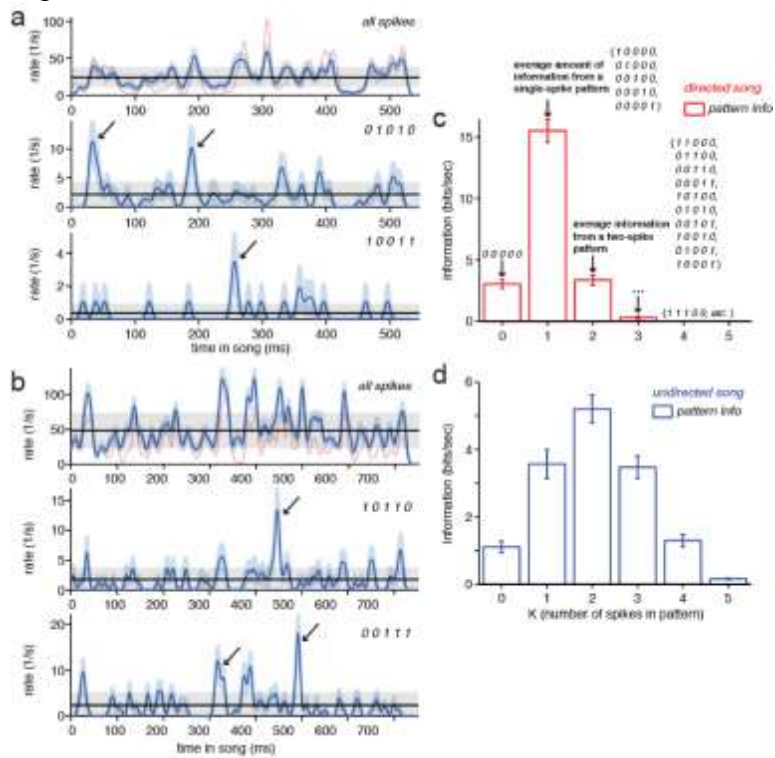

**Figure 3: Bursts carry information about time in song during undirected singing and are, therefore, not purely a noise signal**. (**a**) and (**b**) Smoothed, average firing rates for two different LMAN neurons for single spikes (top) during undirected trials, or particular temporal patterns (middle and bottom) with two or three spikes in a 10ms window. The black line indicates the mean firing rate and the gray area is ± 1 SD from the mean. PSTHs for the same neuron during directed singing are plotted in red for comparison. Peaks in the pattern PSTHs are indicated with black arrows. The data in **a** are from the same site as that in **Figure 1b**. (**c**) Pattern information grouped by count for directed activity. The majority of pattern information comes from single spike patterns. (**d**) Same as **c**, but for undirected activity. During undirected singing, pattern information arises predominantly from 1-, 2- and 3-spike patterns.

Figure 4

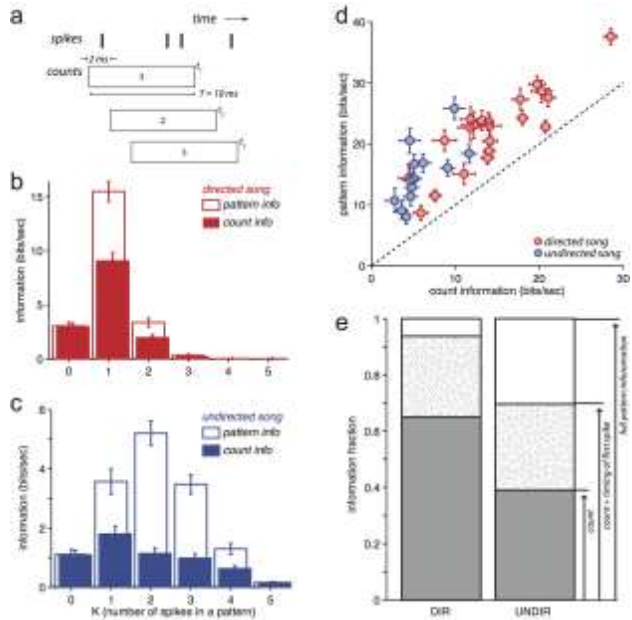

**Figure 4: Temporal coding is more prevalent in undirected song, whereas coding is more rate-like in directed song.** (**a**) An illustration of how we define count and the first spike in the bin for the 10ms windows. (**b**) and (**c**) Different spike patterns carry information in the two singing conditions. Information in patterns (solid lines) versus counts (dashed lines) is plotted versus spike count. During directed singing (**b**), most of the pattern information is contained in single spike events. About half of this information is captured by counts. In contrast, most of the information during undirected singing is carried by higher spike count patterns, and very little information is captured by counts alone (**c**). Error bars indicate SEM across recording sites. (**d**) Information from patterns plotted versus information from count for each recording site in LMAN, in both behavioral conditions. The dashed line indicates perfect rate coding, where pattern info is equal to count info. In undirected song, spike counts carry very little information, though pattern information can be as high as 0.9 bits. During directed song, pattern information is slightly larger than count information, but points follow the rate coding line. Error bars indicate +/- 1 SD. (**e**) Summary of the contributions to total pattern information from count alone (hashed portion) and from count + first-spike-in-the-bin timing (gray portion) reveals a substantial remaining amount of pure temporal sequence information (white portion) during undirected song.

Figure 5

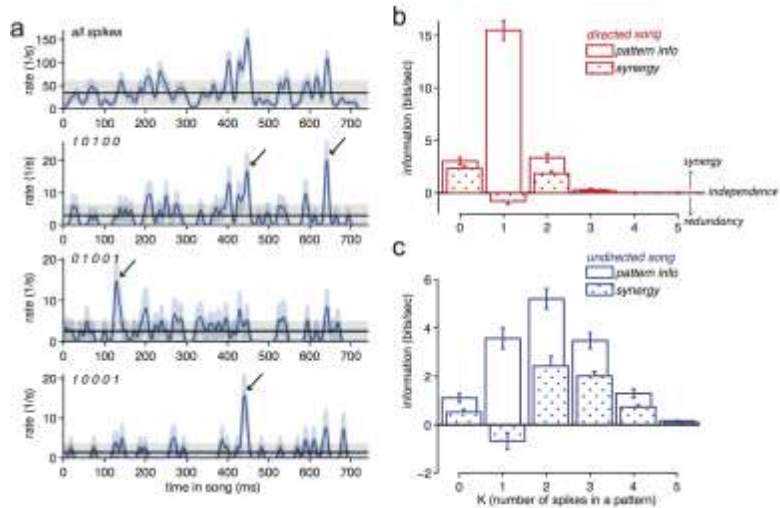

**Figure 5: Bursts code for time in song synergistically.** (**a**) Examples of synergistic coding in 3-spike patterns from a single LMAN neuron. These three distinct patterns peak in probability of occurrence at different parts of song.  The timing of subsequent spikes in the burst determines location in song.  (**b**) Synergy (dotted bars) is plotted for combinations of spikes and silence versus spike count for directed singing as in Eq.4.  Pattern information (open bars) is shown for comparison in the background.  Synergy from silence has a large effect in zero-spike patterns during directed singing.  (**c**) During undirected singing, we observe more synergy in higher-count patterns, corresponding to bursts.

Figure 6

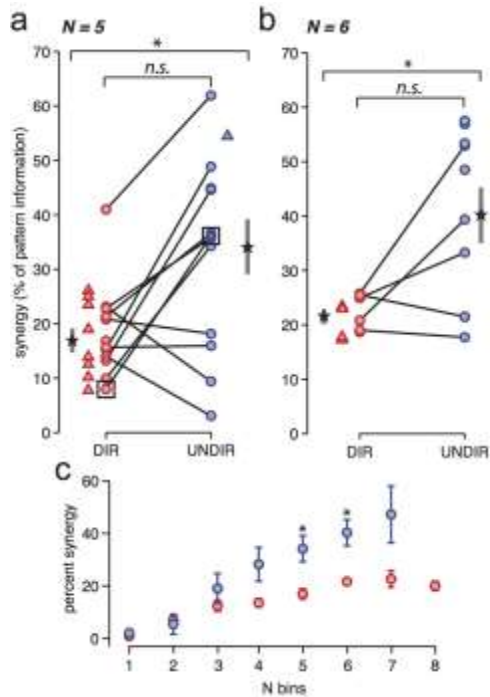

**Figure 6: Overall synergy is higher in undirected song.** (**a**) and (**b**) Averaging over all patterns, we plot the percent of pattern information that comes from synergy for N bins = 5 in **a** and N bins = 6 in **b**. Synergy is significantly higher during undirected singing when assessed by averaging across all sites, both paired and unpaired. The synergy for the site shown in **Figure 5a** is indicated by the black boxes. **c**) The percent synergy also increases with pattern length, and more substantially for undirected versus directed spiking. P-values indicated by asterisks as in **Figure 1**.

# Supplementary Information

Figure S1

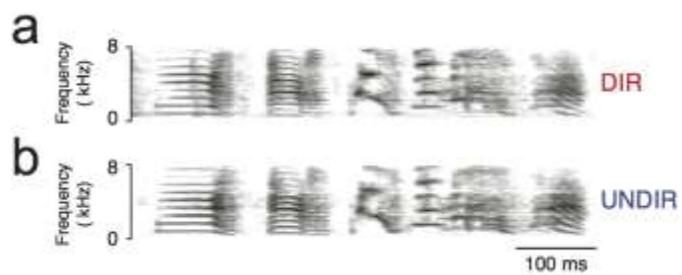

**Figure S1: Song spectrograms are similar in the two behavioral contexts.** (**a**) The median song taken from directed bouts plotted as frequency intensity (ranging from 0 to 8 kHz) versus time in song. (**b**) The median song from undirected bouts.

Figure S2

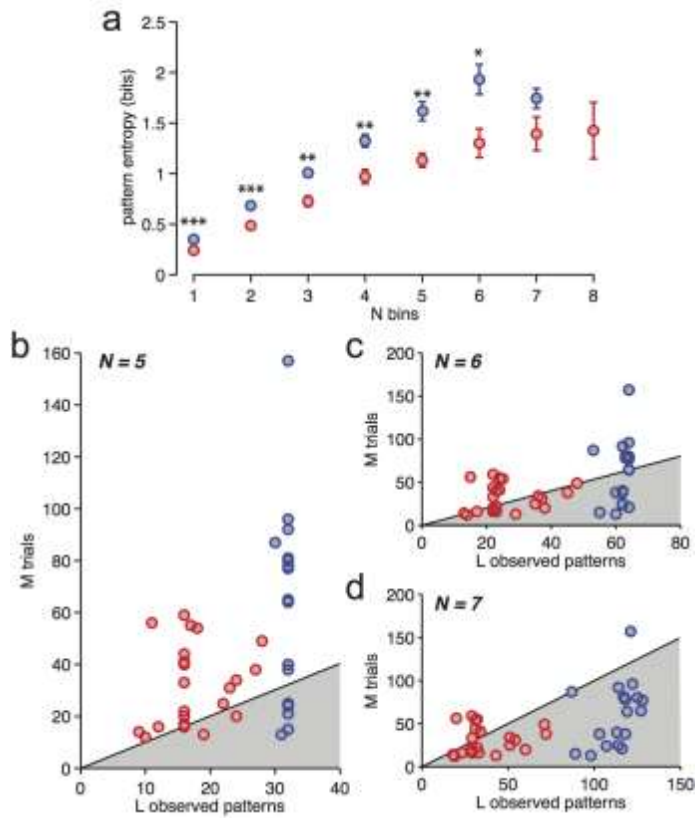

**Figure S2: Pattern entropy increases and number of sites used decreases with increasing N bins.** (**a**) Entropy of patterns of spiking and silence as a function of N bins for directed (red) and undirected (blue) conditions. (**b-d**) The number, M, of trials recorded for each site in each condition versus the number, L, of observed patterns of spiking and silence. This number, L, is always higher during undirected than during directed singing. Sites with fewer trials than observed patterns (gray area) were excluded from our analysis. As N bins increases from 5, shown in **b**, to 6 (**c)** or 7 (**d**), the number of sites we use decreases.

Figure S3

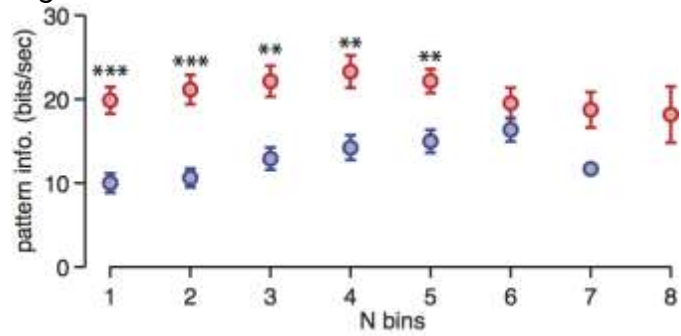

**Figure S3: Pattern information during undirected singing gradually approaches that measured during directed singing.** The average information from temporal patterns over all sites included in this analysis (see **Figure S2**) is plotted versus N bins. Error bars indicated ± 1 SEM and asterisks indicate significance as in **Figure 1**. For N of 7 or 8, we have few sites to sample from and so estimates are noisy.

Figure S4

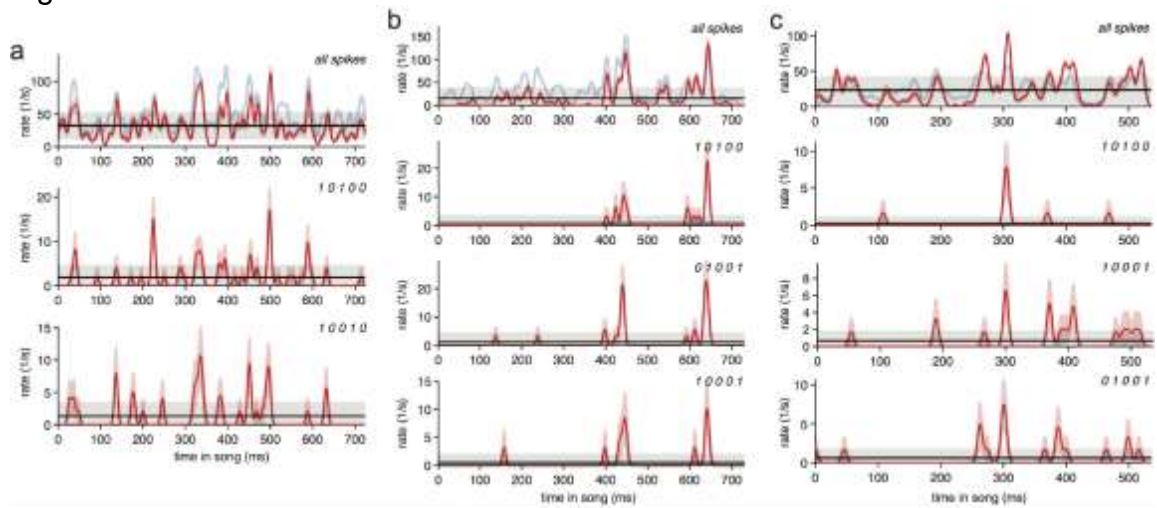

**Figure S4: Directed patterns indicate different parts of song occasionally, but more often code for the same part of song redundantly.** (**a**) Spike or pattern rate as a function of time in song. The gray area indicates ± 1 SD of the rate. The light blue trace is the average spike rate during undirected singing. Patterns 10100 and 10010 point to somewhat non-overlapping parts of song in this site. (**b**) Spike and pattern rates for a different site, showing that these patterns all point to the same parts of song. (**c**) Spike and pattern rates during directed singing for a third site.